\documentclass[multphys,vecphys]{svmult}

\usepackage{makeidx}         
\usepackage{graphicx}        
\usepackage{multicol}        
\usepackage[bottom]{footmisc}

\makeindex             

\begin{document}

\title*{HST observations of gravitationally lensed QSOs}


\author{Jean-Fran\c cois Claeskens\inst{1}, Dominique Sluse\inst{2}
\and
Jean Surdej\inst{1}}


\institute{Institut d'Astrophysique et de G\'eophysique, Universit\'e de Li\`ege, All\'ee du 6 Ao\^ut, 17, B5C, 4000 Li\`ege, Belgium
\texttt{claesken@astro.ulg.ac.be}
\texttt{surdej@astro.ulg.ac.be}
\and 
Laboratoire d'Astrophysique, Ecole Polytechnique F\'ed\'erale de Lausanne (EPFL) Observatoire, CH-1290 Sauverny, Switzerland
\texttt{dominique.sluse@epfl.ch}}
%
%
\maketitle

\begin{abstract}

Thanks to its sharp view, HST has significantly improved our knowledge of tens of gravitationally lensed quasars in four different respects: (1) confirming their lensed nature; (2) detecting the lensing galaxy responsible for the image splitting; (3) improving the astrometric accuracy on the positions of the unresolved QSO images and of the lens; (4) resolving  {\em extended} lensed structures from the QSO hosts into faint NIR or  optical rings or arcs. These observations have helped to break some degeneracies on the lens potential, to probe the galaxy evolution and to reconstruct the true shape of the QSO host with an increased angular resolution.

\end{abstract}

\section{Introduction}
\label{sec:intro}

Strong lensing, i.e. the splitting of the image of a background source into several, magnified but distorted images, occurs each time an intervening massive object lies nearly on the same line-of-sight, provided its surface mass density is large enough (typically larger than 0.5 gr/cm$^2$ at cosmological distances). The angular resolution of the telescope must be sufficient to resolve the images ($\Delta \theta \simeq 1'' \sqrt{M_E/10^{11}M_\odot}$). This is a consequence of the curvature of space-time around massive objects, as predicted by General Relativity. 

\begin{figure}[htb]
\hbox to \columnwidth{

\includegraphics[width=2.8cm]{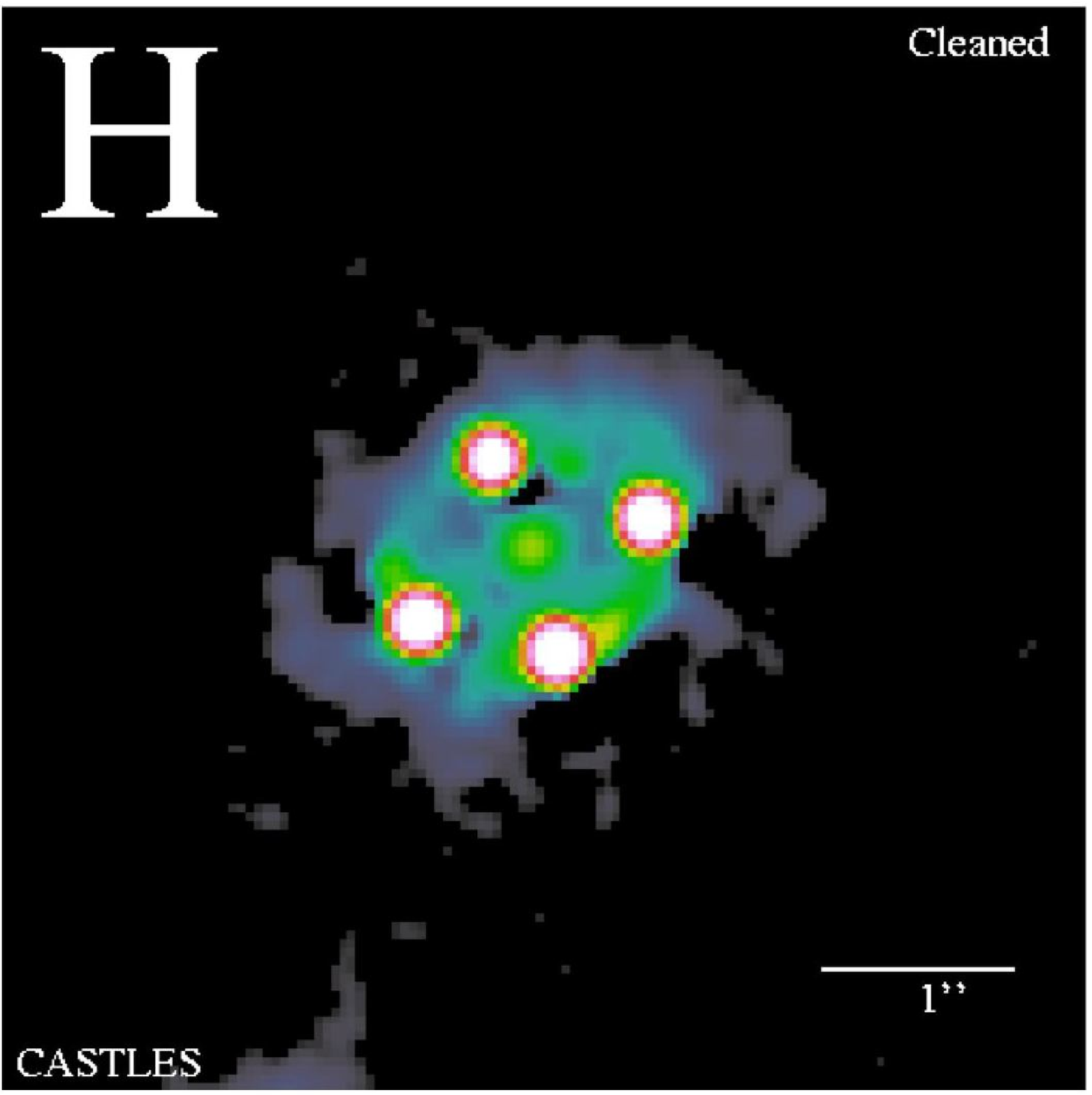}
\includegraphics[width=2.8cm]{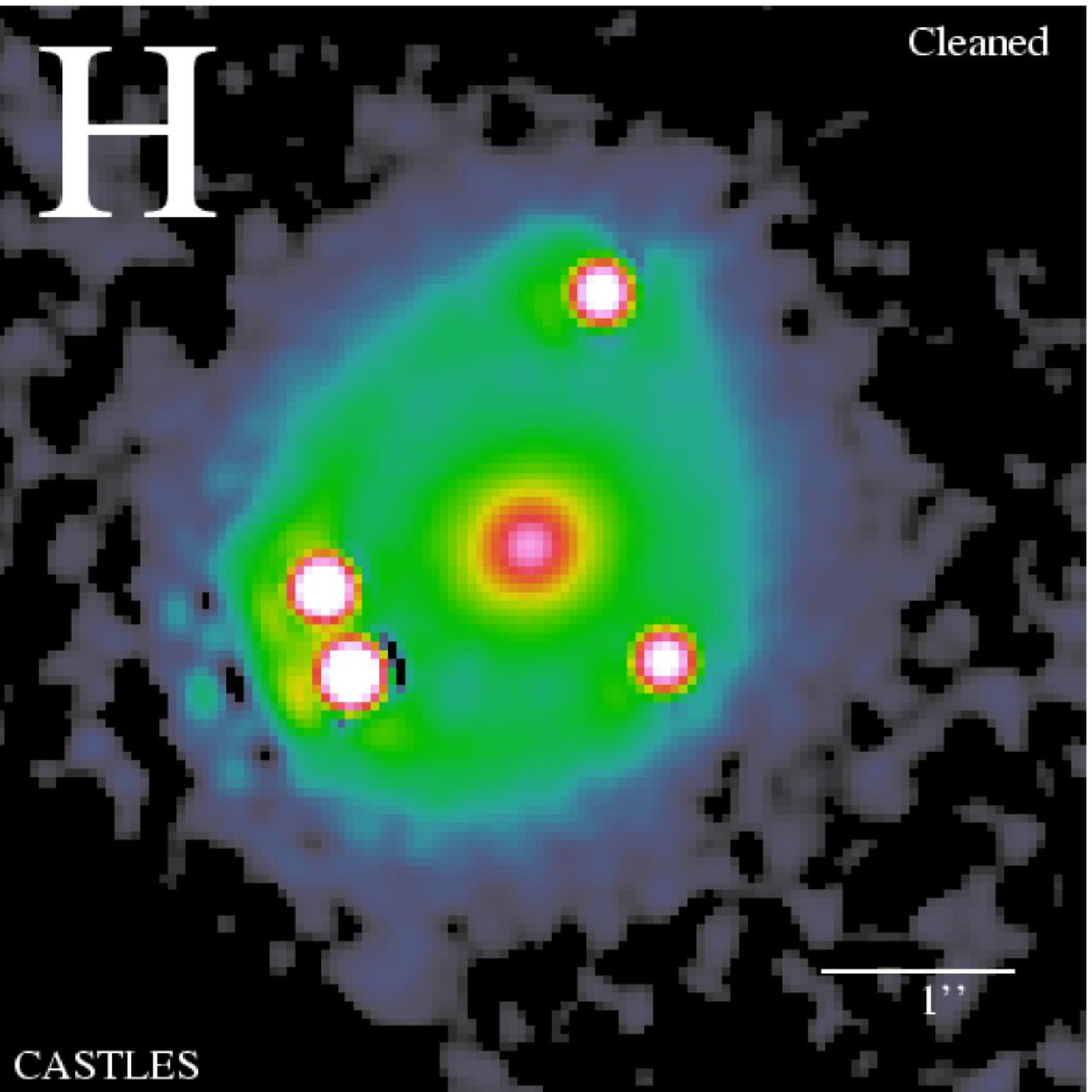}
\includegraphics[width=2.8cm]{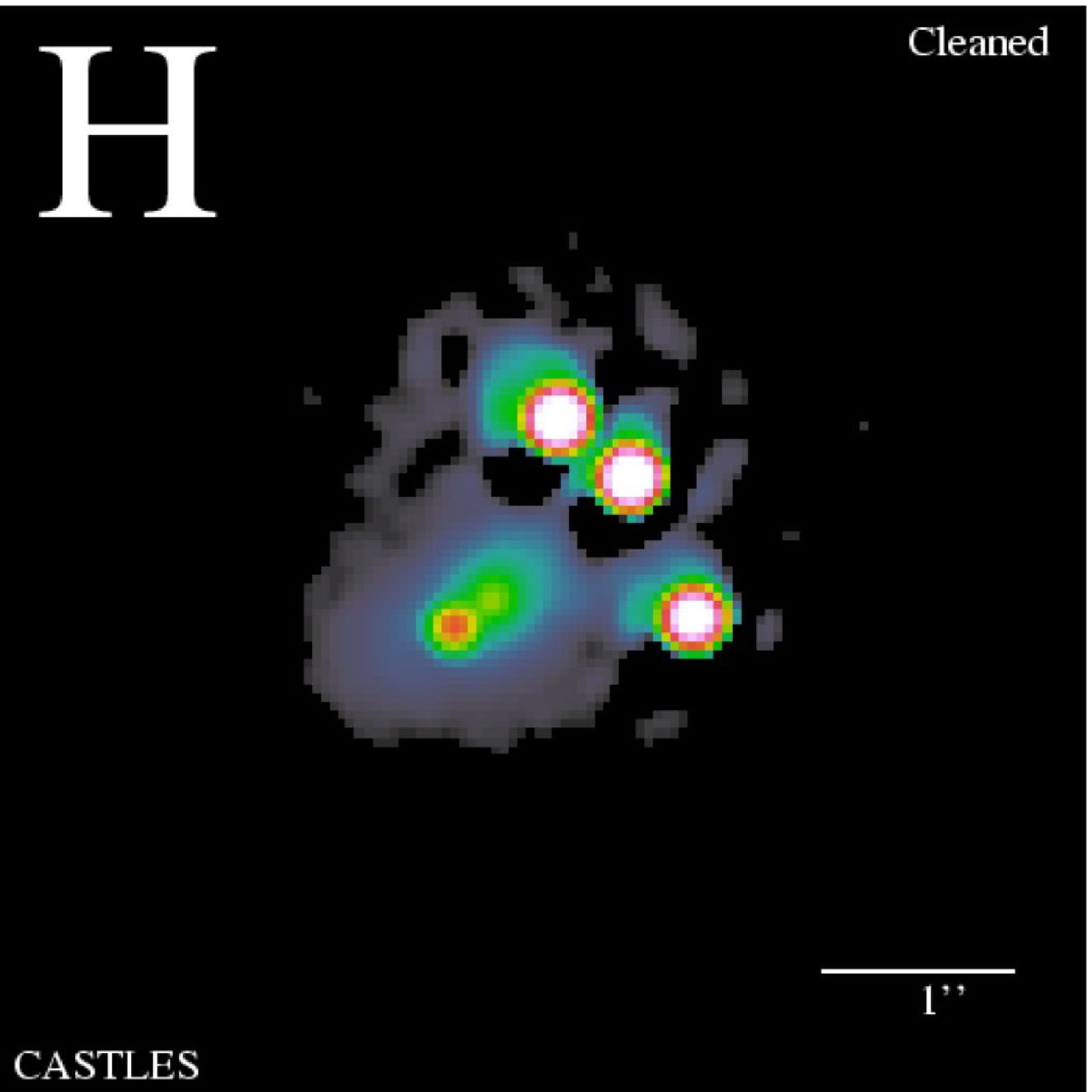}
\includegraphics[width=2.8cm]{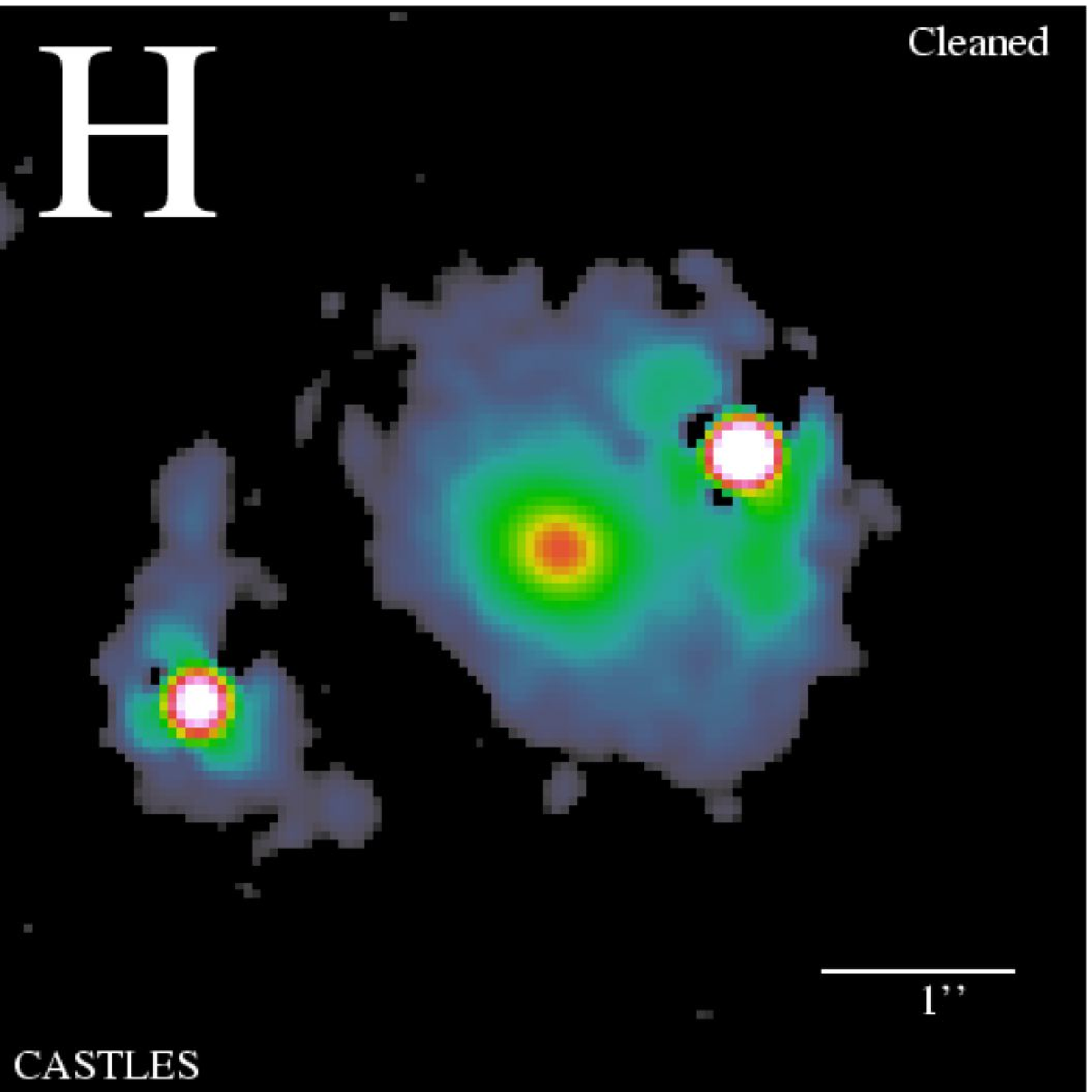}
}

\hbox to \columnwidth{
\includegraphics[width=2.8cm]{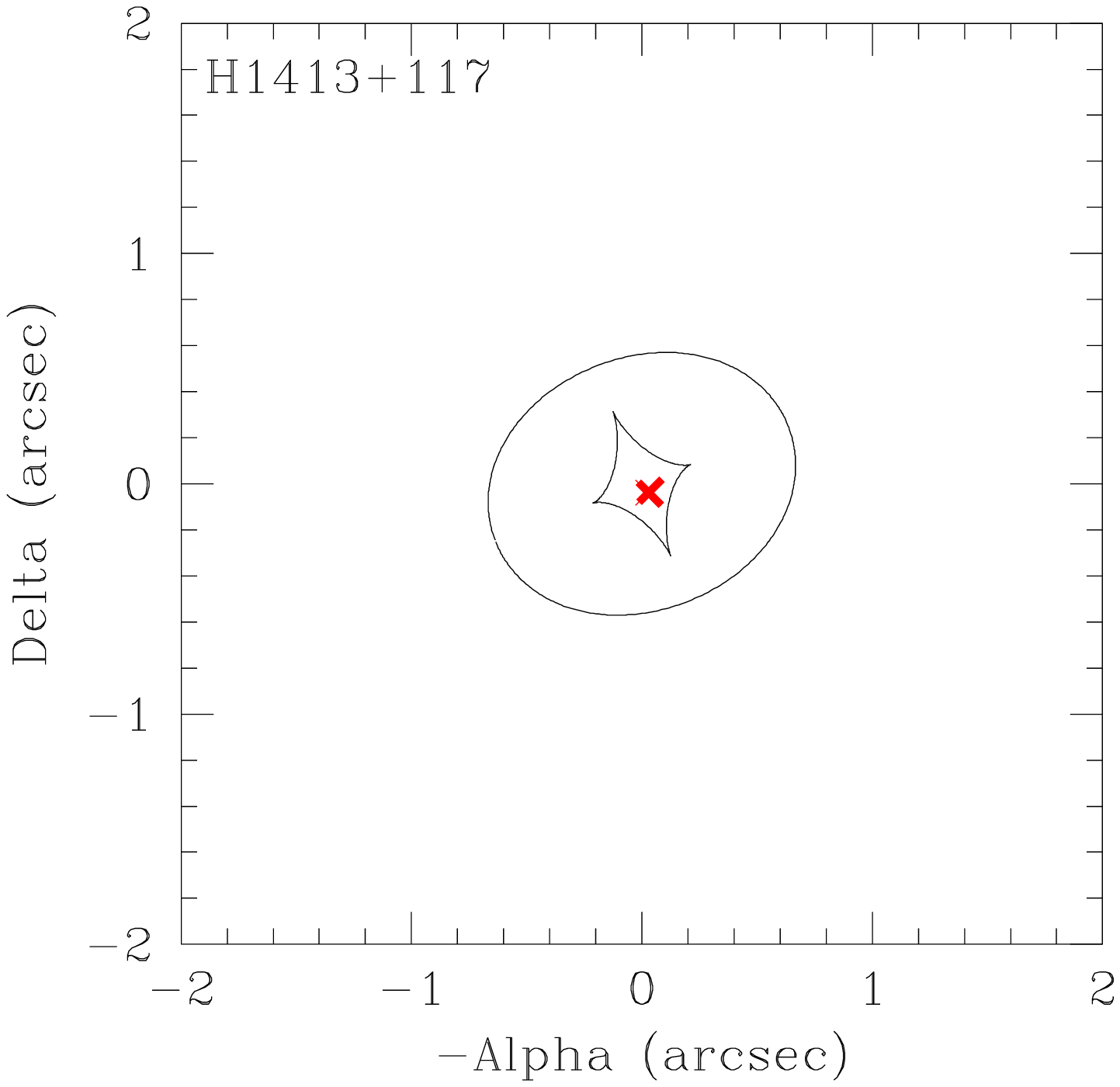}
\includegraphics[width=2.8cm]{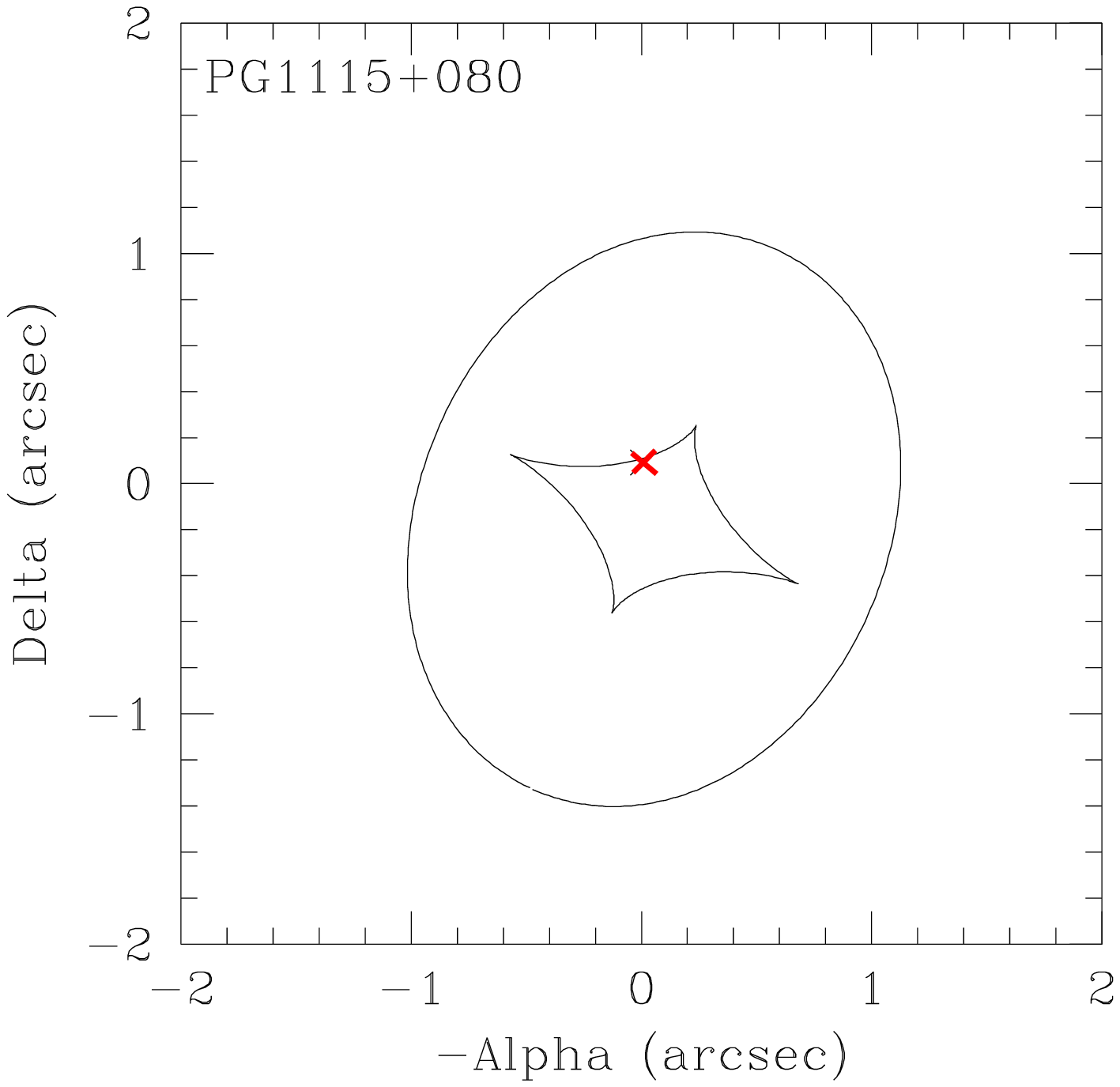}
\includegraphics[width=2.8cm]{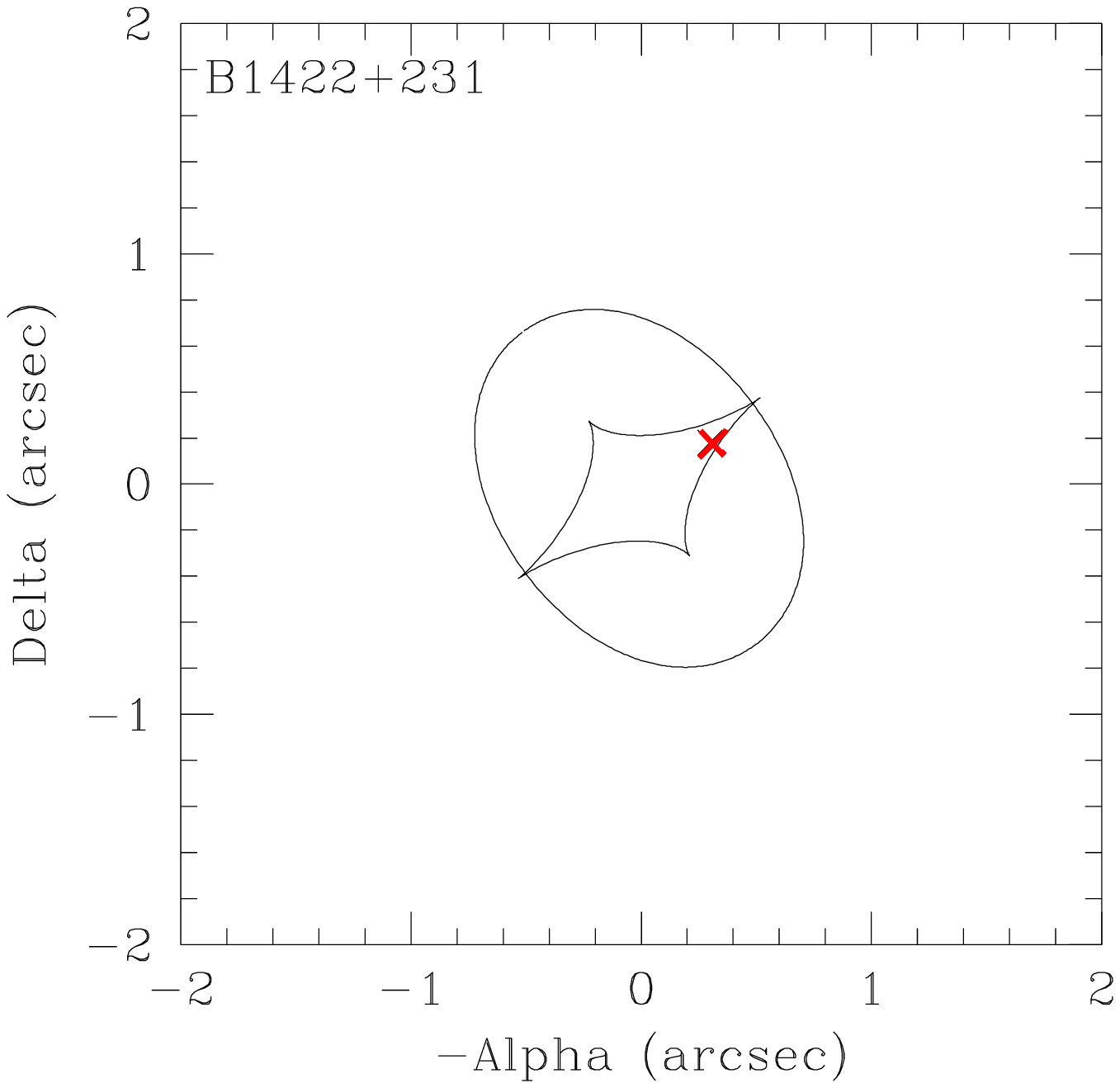}
\includegraphics[width=2.8cm]{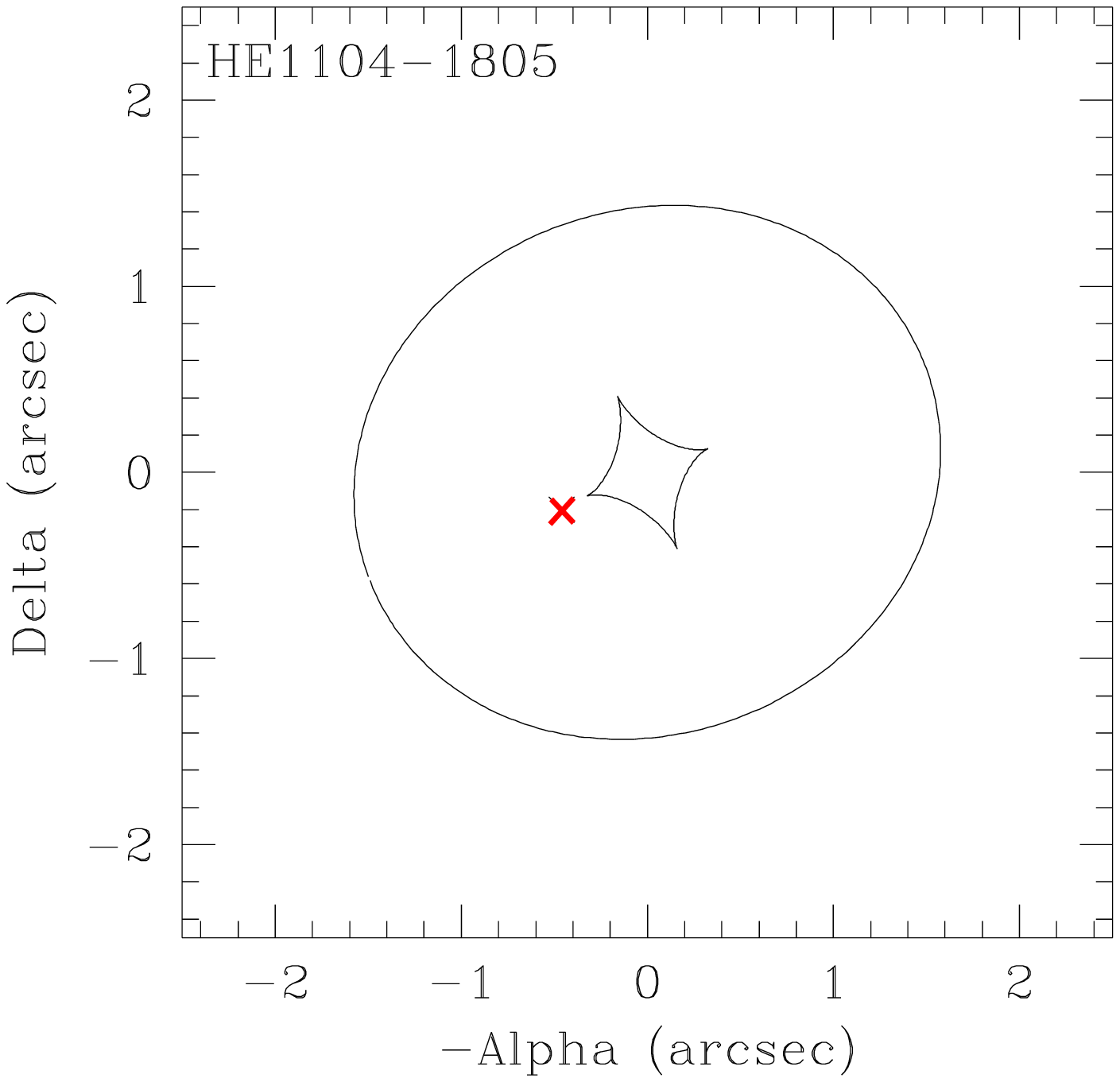}
}
\caption{{\em Top:} Deconvolved HST NICMOS observations of 4 gravitationally lensed QSOs (CASTLES project, {\sl http://cfa-www.harvard.edu/glensdata}); {\em bottom:} fitted source position w.r.t. the caustics of the Singular Isothermal Ellispoid model. From left to right: 4 image cross, fold and  cusp configurations, 2 image configuration.}
\label{fig:lens_config}
\end{figure}

However, a massive object such as an isolated galaxy can be considered as an imperfect optical lens (like the foot of a glass of wine). Instead of a focus, a generic diamond-shaped closed curve is produced -- the caustic. All observed configurations of multiply imaged QSOs as well as giant arcs can be explained by the exact relative position of the (possibly extended) source with respect to that caustic (see Fig. \ref{fig:lens_config}). Of course, besides being an exotic curiosity, this phenomenon betrays the nature of the deflector and can be used as an astrophysical tool. Reviews on gravitational optics and its astrophysical applications may be found in \cite{rev06} and references therein. Here, we concentrate on selected results derived from HST observations of lensed QSOs.

In 1979, the first gravitationally lensed source was discovered serendipitously. The source was the radio-loud QSO Q0957+561A\&B. However, as demonstrated later\cite{rat99} with HST, only relying on {\em chance} to discover lensed sources leads to preferentially select {\em faint} ones, which are then more difficult to follow up and to characterize and which are further embedded in complicated observational biases. 

However, as first mentioned in the 80s\cite{tog84}, {\em luminous} quasars are excellent candidates to look for multiple images produced by isolated galaxies: the optical depth for lensing is high towards those distant sources {\em and} their luminosity function is steep, introducing an amplification bias, which boosts the probability of lensing by a factor of 10 or more. Although about 100 such systems are presently known, HST has only contributed to one discovery through a QSO survey (i.e. Q1208+1011\cite{bah92})! Indeed, ground based optical and radio QSO surveys (and recently the SDSS survey) were much more efficient and faster. But starting from here, HST has played a crucial role.

\section{Confirming the lensing nature}

While a set of 4 QSO images is conspicuous of lensing, double image configurations are not a signature unique to lensing. In the case of small angular separations, spectroscopy with the Faint Object Spectrograph onboard HST was required to assess the true statistics of lensing in optical samples (i.e. Q1208+1011A\&B\cite{bah92} \& J03.13A\&B\cite{sur97}).

\section{Improving the constraints on the lens model}

The generic properties of the lensing potential can be described with  7 parameters: its center position with respect to the lensed images ($\theta_x,\theta_y$), the total mass within the images $M_E$, the total radial mass profile (logarithmic slope $\beta$, core radius $r_c$) and the shape of the iso-density contours ($\epsilon, \varphi$). Moreover, the influence of the direct environment of the lens (neighbours, cluster,...) may be described by 3 extra parameters: an external shear $\gamma$ with orientation $\varphi_\gamma$ and a possibly associated matter density $\kappa_c$. 

However, the insight we can really get into the lensing potential is related, firstly, to the number of observational constraints, secondly to the accuracy on each of them, and finally to the existence of internal degeneracies between different properties of the potential.

\vspace{0.25cm}
\noindent
{\bf The lens position}

\vspace{0.25cm}
Finding the faint distant galaxy in the vicinity of the bright QSO images is a challenge. HST high angular resolution and infra-red capabilities (NICMOS) have provided the direct detection, the precise astrometry and sometimes the morphology of many lenses (e.g. HE1104-1805\cite{rem98}, Q1009+025\cite{cla01}, PG1115+080\cite{imp98}), thus increasing the number of constraints in these systems.

\begin{figure}[b]
\centerline{\includegraphics[width=6.cm]{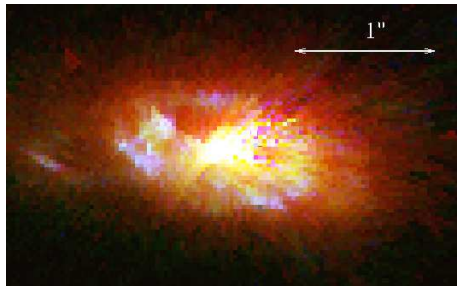}}
\caption{Reconstructed shape of the lensed Seyfert I galaxy RXS J1131-1231\cite{cla06}}
\label{fig:1131}
\end{figure}

\vspace{0.25cm}
\noindent
{\bf The image astrometry}

\vspace{0.25cm}
Reaching an astrometric accuracy of $\sim$ 3 mas, HST observations of gravitationnally lensed quasars have shown that simple lens models (such as the Singular Isothermal Ellipsoid) systematically fail in reproducing the exact image and lens positions. Adding an external shear drastically improves the fit\cite{kee97}, independently of the adopted exact radial profile (e.g. slope, core radius\cite{wit97}). The origin of this external shear is not entirely clear. While in some cases, the external shear derived from the environment can explain the lensed image positions (e.g. in PG1115+080\cite{sch97}), it is  usually not strong enough\cite{fau04}.

\vspace{0.25cm}
\noindent
{\bf Extended structures}

\vspace{0.25cm}

Because the source position is unknown, the number of astrometric constraints is $2(N-1)$ in a $N$ image system. This is just sufficient to determine $M_E, \epsilon, \varphi, \gamma$ and $\varphi_\gamma$, if $N=4$. However, a degeneracy is observed between $\epsilon, \gamma$ and $\varphi - \varphi_\gamma$, whose strength depends on the exact lens configuration\cite{kee97}. There is also a degeneracy between  $\kappa_c$ and $M_E$ or between $\kappa_c$ and $\beta$\cite{wuc01}.

The detection of {\em resolved} arcs and ring-like features (i.e. [part of] the lensed QSO host) may help in breaking some of these degeneracies if they provide extra azimuthal and/or radial constraints. Such detections are one of the best achievements of HST. The first ones were made in the NIR (e.g. PG1115+080\cite{imp98},...) and have been recently supplemented by optical rings, such as RXS J1131-1231\cite{cla06}.

\section{Investigating the nature of the lens and the source}

Besides numerous studies of the individual lensing systems, the above mentioned HST observations have allowed {\em statistical} analysis of the lens population, which represents a {\em mass}-selected sample in low density environments. An interesting example is the evolution of their $M/L$ as a function of redshift through the Fundamental Plane analysis: it is found to be comparable to that observed in galaxy clusters\cite{vandeven03}. On the other hand, a moderate but measurable amount of patchy extinction in the lenses has also been reported\cite{fal99}.

Finally, the HST high angular resolution combined with the lens magnification is opening deeper insight into the QSO {\em host galaxy}. As an example, Fig. \ref{fig:1131} shows the reconstructed image of a Seyfert I host galaxy, where spiral arms, star-forming regions and interaction with a companion are clearly seen at a redshift $z=0.66$. On the other hand, a {\em statistical} study of the lensed host galaxies has shown the rise of $M_{\rm bh}/M_*$ with redshift as far as $z\sim 4$\cite{pen06}. 

\vspace{0.1cm}
\noindent
{\em Acknowledgement:} This work was supported by PRODEX PEA C90194 HST.

%
%
%



\printindex
\end{document}